\documentclass{emulateapj}

\newcommand{\Te} {T_{\rm eff}}
\newcommand{\logg} {\log g}

\begin{document}

\title{SDSS~J142625.71+575218.3: The First Pulsating White Dwarf with a 
Large Detectable Magnetic Field} 

\author{P. Dufour\altaffilmark{1},
G. Fontaine\altaffilmark{2},
James Liebert\altaffilmark{1},
Kurtis Williams\altaffilmark{3,4},
David K. Lai\altaffilmark{5},
}

\altaffiltext{1}{Steward Observatory, University of Arizona, 933 North
  Cherry Avenue, Tucson, AZ 85721; dufourpa@as.arizona.edu,
  liebert@as.arizona.edu}

\altaffiltext{2}{D\'{e}partement de Physique, Universit\'{e} de
  Montr\'{e}al, C.P. 6128, Succ. Centre-Ville, Montr\'{e}al,
  Qu\'{e}bec, Canada H3C 3J7; fontaine@astro.umontreal.ca}
 
\altaffiltext{3}{Department of Astronomy, University of Texas at
  Austin, Austin, TX, USA; kurtis@astro.as.utexas.edu}

\altaffiltext{4}{NSF Astronomy and Astrophysics Postdoctoral Fellow}

\altaffiltext{5}{Department of Astronomy and Astrophysics, University
  of California, Santa Cruz, CA 95064; dklai@ucsc.edu}

\begin{abstract}
We report the discovery of a strong magnetic field in the unique
pulsating carbon-atmosphere white dwarf SDSS J142625.71+575218.3. From
spectra gathered at the MMT and Keck telescopes, we infer a surface
field of $B_s$ $\simeq$ 1.2 MG, based on obvious Zeeman components
seen in several carbon lines. We also detect the presence of
a Zeeman-splitted He I 4471 line, which is an indicator of the 
presence of a non-negligible amount of helium in the atmosphere of this
Hot DQ star. This is important for understanding its pulsations, as
nonadabatic theory reveals that some helium must be present in the envelope
mixture for pulsation modes to be excited in the range of effective
temperature where the target star is found. Out of nearly 200 pulsating
white dwarfs known today, this is the first example of a star with a
large detectable magnetic field. We suggest that SDSS
J142625.71+575218.3 is the white dwarf equivalent of a roAp star.
\end{abstract}


\section{INTRODUCTION}

Recently, \citet{dufournat} reported the discovery of a new type of
white dwarf stars with an atmosphere composed primarily of carbon with
little or no traces of hydrogen or helium (the ''Hot DQ'' spectral
type). Prior to that discovery, white dwarfs cooler than $\sim $ 80,000 K
were known to come in essentially two flavors: those with an almost pure
hydrogen surface composition (forming the DA spectral type), and those
with a helium-dominated surface composition (the non-DA stars, which
comprise the DO, DB, DC, DZ, and DQ spectral types). Pulsationally
unstable stars are found among these two broad families of white dwarfs,
in each case occupying a narrow range of effective temperature.
Variable white dwarfs situated in these instability strips are,
respectively, classified ZZ Ceti stars (hydrogen atmospheres, $\Te \sim$
12,000 K) and V777 Her stars (helium atmospheres, $\Te \sim$ 25,000
K). Since each of these instability regions is associated with 
the presence of a partial ionization zone of the primary atmospheric
constituent (H or He), it was naturally expected that some 
carbon-atmosphere white dwarfs could be unstable as well in a 
certain regime of effective temperature corresponding to partial
ionization of carbon. And indeed, \citet{fontaine76}  found
long ago strong similarities between the partial ionization regions and
associated superficial convection zones of white dwarf models with H-,
He-, and C-dominated atmospheres/envelopes. Theoretical considerations
thus suggested that some of the Hot DQ white dwarfs, because they are
located in a narrow range of effective temperature around 20,000 K
\citep{dufournat,dufour08}, could possibly pulsate. Following this, a systematic
search for pulsations in carbon-atmosphere white dwarfs carried out by
\citet{montgomery08} successfully discovered the first pulsating
carbon-dominated atmosphere white dwarf: SDSS~J142625.71+575218.3
(hereafter SDSS~J1426+5752). They uncovered a single pulsation (and its
first harmonic) with a period of 417 s in that star. 

In parallel with this observational effort, \citet{fontaine08} carried
out the first detailed stability study, based on the full nonadiabatic
approach, to investigate the asteroseismological potential of
carbon-atmosphere white dwarfs. They showed that pulsational
instabilities in the range of effective temperature where the Hot DQs
are found are indeed possible, but only if a fair amount of helium is
present in the atmosphere/envelope compositional mixture. White dwarf
models with pure carbon envelopes are found to pulsate, but only at
much higher temperatures than those characterizing Hot DQ's discovered
up to now. However, the SDSS spectra analyzed in \citet{dufour08},
even though quite noisy, rule out large amounts of helium (from the
absence of the He~\textsc{i} $\lambda$4471 line) for all objects
except SDSS~J1426+5752 which, because of its higher surface gravity
and lower effective temperature, could perhaps have an He/C abundance
ratio as high as 0.5. So, according to the full nonadiabatic models,
SDSS~J1426+5752 should be the only object pulsating, as observed,
provided helium is present in a relatively large amount at the
surface. However, even with such an abundance (He/C = 0.5), only a
tiny depression at the He~\textsc{i} $\lambda$4471 line is predicted
by synthetic models. Given the noisy SDSS spectrum for this object, no
firm conclusion could be reached concerning the presence of helium,
although the above mentioned abundance looked quite compatible with
the spectroscopic observation (see the SDSS spectrum and fits in
Figure 1).

All this motivated us to obtain higher sensitivity observations for
this special object in order (1) to confirm, as may be the case, the
presence of a relatively large amount of helium (thus increasing our
confidence in the nonadiabatic approach), and (2) to be able to
eventually carry out a full asteroseismological analysis using better
constraints/determinations of the atmospheric parameters ($\Te$,
$\logg$ and He/C) from spectroscopy. In this Letter, we thus report
new high signal-to-noise ratio spectroscopic observations (MMT and
Keck) that revealed, to our great surprise, that SDSS~J1426+5752 is
the first pulsating white dwarf showing clear evidence for the
presence of a strong magnetic field. This is certainly an unexpected
result, given that strong magnetism in white dwarfs is generally
thought to extinguish, or at least greatly diminish, pulsational
activity. We discuss below the implications of this discovery.

\section{OBSERVATIONS}\label{observation}

Since the signal-to-noise ratio for the SDSS spectra of the faint
known carbon-atmosphere white dwarfs is not sufficient for a precise
determination of the atmospheric parameters, a program to reobserve
all the Hot DQ stars with the MMT 6.5 m telescope was recently
undertaken. The complete analysis of these new Hot DQ spectra will be
presented in due time, once the program is completed. As a part of the
program, SDSS~J1426+5752 was observed for a total of 180 minutes with
the Blue Channel spectrograph on the night of 2008 May 5. We used the
500 line mm$^{-1}$ grating with a 1'' slit, resulting in a $\sim$~3.6
\AA~FWHM spectral resolution over a wavelength range of 3400-6300
\AA. The spectra were reduced with standard IRAF packages. The final
combined spectrum, shown in Figure 1, has a signal-to-noise ratio of
$\sim$75 at 4500 \AA.

A lower signal-to-noise observation was obtained at the Keck
Observatory on the night of UT 2008 May 4 with the blue channel of
LRIS. The 400 groove mm$^{-1}$, 3400\AA~ grism was used with the D560
dichoric and the atmospheric dispersion corrector was active.  Two
exposures with a total exposure time of 1260 seconds were taken with a
0\farcs7 slit in 0\farcs9 seeing at an airmass of 2.3. The spectra
were reduced and extracted using standard IRAF packages. The spectra
were corrected for atmospheric extinction using the IRAF KPNO
extinction curve; the measured signal-to-noise is $\approx 54$ per
5\AA~ resolution element at 4500\AA.

In Figure 1, we present our new high signal-to-noise ratio MMT
spectrum, the Keck observation, as well as the SDSS
data that were used for the \citet{dufour08} analysis. The most
striking revelation brought by these new observations is that many of
the carbon features are clearly well separated into three Zeeman
components (see bottom panels in Figure 1). The separation between the
components of the C~\textsc{ii} features corresponds to a surface field
$B_s \approx$ 1.2 MG. Note that with the poor signal-to-noise
ratio of the SDSS observation, \citet{dufour08} could not resolve
the Zeeman structure and, as a result, their spectroscopic solution (see
Fig. 1) can now only be considered as an approximation of the true atmospheric
parameters. Our new observations also reveal a small but quite
significant depression near the 4471 He~\textsc{i} line, indicating that
helium is indeed present in relatively large abundance in this
object. However, since our atmosphere models do not include a magnetic
field yet, no exact abundance can be derived at this point, although it
is probably not too far from the 0.5 value mentioned above.

\begin{figure}[h]
\plotone{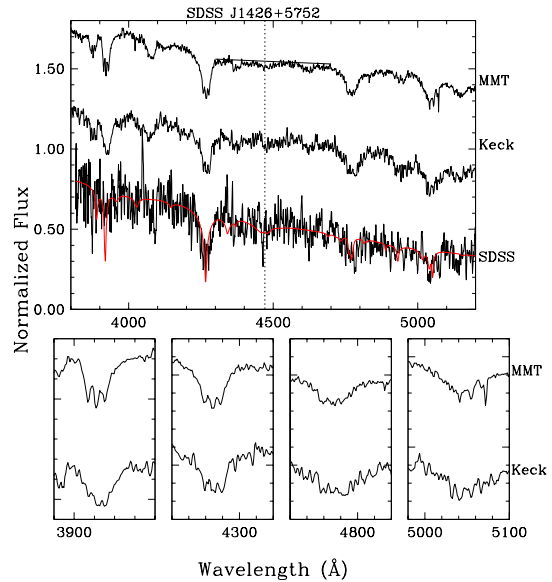}
\caption{{\it Top:} Spectroscopic observations from SDSS, Keck and
MMT, normalized to unity at 4500 \AA~ and offset from each other for
clarity. In red is the solution of \citet{dufour08} based on the
nonmagnetic fit of the noisy SDSS spectrum ($\logg$ = 9.0, $\Te$ =
19,830 K and He/C = 0.5). The dotted line marks the position of the
He~\textsc{i} $\lambda 4471$ line. We also added, to show the presence
of the small helium absorption, a line representing approximatively
the continuum for the MMT spectrum (the other depression near 4625
\AA~ is a carbon feature usually observed in higher $\Te$ objects,
indicating that the solution of \citet{dufour08} is perhaps slightly
underestimated). Note that the SDSS and Keck spectra have been
smoothed with a 3 pts average window for clarity {\it Bottom:}
Selected carbon lines from the Keck and MMT spectra that demonstrate
the presence of a 1.2 MG magnetic field through Zeeman splitting.}
\end{figure}

\section{DISCUSSION}\label{results}

\subsection{Origin of the Magnetic Field}

It is generally believed that many magnetic white dwarfs are the
likely descendants of magnetic main sequence stars (the Ap and Bp
stars), and that the high fields observed in some white dwarfs are the
result of (partial) magnetic flux conservation of a fossil field as
the star shrinks in radius by a factor of $\sim$ 100 as it becomes a
white dwarf. Since, however, about ten per cent of isolated white
dwarfs are high field magnetic white dwarfs \citep[recognizable from
Zeeman split spectral lines,][]{liebert03,kawka07}, another origin may
also be necessary, possibly as proposed by \citet{tout08}. 

Dynamo-type mechanisms have also been proposed to explain the putative
weak magnetic field in the pulsating DB white dwarf GD 358
\citep{markiel94,thomas95}. However, it is quite unlikely that such
mechanisms could account for a field as high as 1.2 MG in the
carbon-rich atmosphere of SDSS~J1426+5752, or in a white dwarf in
general. Indeed, to produce a dynamo-type magnetic field, convective
motions must be strong enough to twist and move seed magnetic
lines. As the field grows, due to dynamo amplification, convection is
having a harder time to move the field lines. Thus, the final
large-scale field can never reach an amplitude comparable to the
so-called equipartition field strength given by the condition
$B_{eq}^2/8\pi = 1/2 ~ <\rho ~ v_{conv}^2>$, where the last term
corresponds to a suitable average over the convection zone of the
convective energy density. 

Typical values of $B_{eq}$, calculated for H-, He- and even
C-atmosphere white dwarfs by \citet{fontaine73} are $\sim$ 10-100
kG. While these results need to be revisited, in particular using a
state-of-the art model for SDSS~J1426+5752, it would be extremely
surprising to find that the order of magnitude estimates of
\citet{fontaine73} could change significantly. We thus believe that
the 1.2 MG magnetic field found in SDSS~J1426+5752 is a fossil field,
probably originating from an Ap star.

\subsection{Magnetism and Pulsations in White Dwarf Stars}

To date, there is no clear evidence for the presence of an
observable magnetic field in any of the known pulsating white
dwarfs. None of the 51 bright ZZ Ceti stars from the Bergeron sample
show any sign of Zeeman splitting in the optical spectra, which
translates, given the typical S/N ratio and resolution of the
observations, to limits on the magnetic field strength of about 500 kG
(P. Bergeron, private communication). Also, none of the known pulsating
DB white dwarfs have a magnetic field strong enough to be detected from
Zeeman splitting. This is also the case for the 18 known pulsating white
dwarfs of the GW Vir type.

In order to detect weaker magnetic fields down to a few kG, 
spectropolarimetric measurements are needed. Unfortunately, only a small
number of pulsating white dwarfs have been investigated with this more
precise method. Nevertheless, no significant magnetic field has ever
been found in any of the few pulsating white dwarfs for which
spectropolarimetric measurements are available
\citep{schmidt97,valyavin06}, and very small upper limits of a few kG
are obtained in all cases (with perhaps an uncertain marginal detection
in one case).

The fact that the samples of magnetic and pulsating white dwarfs do not
intersect may not be very surprising from a theoretical point of
view. Indeed, pulsating white dwarfs of both the V777 Her and ZZ Ceti types
are found in a regime of effective temperature where an important
superficial convection zone is present. The latter is due to the partial
ionization of either He or H, and contributes significantly to the
excitation of pulsation modes. For a large scale magnetic field of
magnitude much stronger than the equipartition field strength, it is
likely that the convective motions are largely quenched, which perhaps
extinguishes completely pulsational driving. One example of a
white dwarf where a magnetic field \citep[$B_e = -1000 \pm
500$ kG,][]{putney97} might have ''killed'' the pulsations is the constant
DB star LB~8827 \citep[PG~0853+164,][]{wesemael01}. Unfortunately, the
effective temperature of this object is uncertain, and it is not
known with certainty whether it is inside the DB instability strip or
not.

The only case where the detection of a magnetic field has been claimed
in a pulsating white dwarf is that of GD~358. In that case, the magnetic
field has been indirectly inferred from asteroseismological analysis
\citep{winget94}. It should be noted that this interpretation of the
asteroseismological data in terms of a magnetic field is far from being
accepted by all (see, e.g., Fontaine \& Brassard 2008, in preparation). In any
case, follow-up circular polarization measurements of GD ~358 by
\citet{schmidt97} have not succeeded in detecting the presence of a
weak field, but their detection threshold was significantly above 
the value of 1300 $\pm$ 300 gauss suggested by \citet{winget94}. Such
a small field is certainly not strong enough to affect the convection zone
significantly, and is apparently unable to stop the pulsations in this
variable white dwarf.

\subsection{Rotation or Pulsations?}

In this section, we briefly discuss the possibility that rotation might
be a significant ingredient in this puzzle. Indeed, rapid rotation is
known to be important in at least two variable magnetic white dwarf
systems where the variability is explained by changes with rotational
phase instead of pulsational instabilities. The first of these cases,
RE~J0317-853 \citep{barstow95, burleigh99}, is a highly magnetic, 
rotating white dwarf with a period of 725 s that is most probably the
result of a double degenerate merger. The second case, Feige 7
\citep{liebert77}, is also a rotating magnetic white dwarf but, this
time, with a period of 2.2 hours. Its spectrum shows Zeeman splitting
for both hydrogen and helium that appears to vary with rotational phase 
\citep{achilleos92}.

Could it be that SDSS~J1426+5752 is a rare magnetic white dwarf
spinning very fast (which would make it, with a period of 417 s, the
fastest white dwarf amongst isolated white dwarfs)? Several factors
lead us to believe that, on the contrary, this star is most likely a
pulsator and not a rotator. The exposure time for each of our
integrations at the MMT (600 s) is well above the period of 417 s
found by \citet{montgomery08}, meaning that our spectra are averaged
over a variability cycle. If the luminosity variations are due to fast
rotation of the star, it is quite probable that the average magnetic
field along our line of sight, depending on the geometry and the
alignment of the field with respect to the rotation axis or the
presence of magnetic spots, would vary over one cycle. The resulting
Zeeman splitting of atomic line should thus vary in magnitude as the
field strength changes, leading to very broad or blended lines in our
average spectra, not three well separated and sharp components as
observed (see bottom panels in Figure 1).

Of course, one could imagine a situation where a dipole field is well
aligned with the rotation axis, or a more complex field geometry that
is such that the field remains almost constant over the rotation
period although our knowledge of other magnetic white dwarfs suggests
that this is quite rare and unlikely \citep{wickramasinghe02}. This,
alone, is not an argument strong enough to discard completely this
possibility of fast rotation. 

However, our own rapid photometry campaign at the 61'' telescope at
Mount Bigelow with the Mont4K CCD imager brings an important new piece
to this puzzle (Fontaine et al. 2008, in preparation). Indeed, based
on a total of 107 hours of observations spread over one month (half in
early April and half in early May 2008), the presence of a
low-amplitude second mode with a period of 319 s (a 4.9 sigma result)
has now been revealed. That SDSS~J1426+5752 is therefore very likely a
multiperiodic pulsator is certainly a strong argument against the fast
rotator hypothesis. A more standard pulsation mechanism, although
involving a strong magnetic field, is thus probably at work here (see
below).

Finally, if we look individually at each combination of two
consecutive 600 s MMT exposures (a single exposure is too noisy
to reveal the splitting in the lines), we do not find any evidence for
a change in the separation of the Zeeman components over a three hour
period, indicating that the magnetic field strength remains constant
over that timescale. Using the Keck data from the previous night,
we find that, within the limits of the noise, the spectrum looks
unchanged on a $\sim$20 h period as well. Unless we are dealing
with a complicated magnetic field geometry or that a dipole field is
perfectly aligned with the rotation axis, this probably indicates that
this star is rotating very slowly, which is more in line with our 
understanding that magnetic white dwarfs are generally slow rotators
\citep{wickramasinghe02}.

\subsection{A roAp Star Analog?}

As discussed above, the magnetic field in SDSS~J1426+5752 is certainly
much stronger than $B_{eq}$, which immediately suggests that the
convective motions are smothered by the field and perhaps even that
there is no convection at all. At the very least, because ionized matter
cannot freely cross field lines, one would expect the suppression of
convective motions in the magnetic equatorial regions while some
``channeled'' motions could resist near the magnetic poles. Since
convection plays a role in the driving of pulsation modes in nonmagnetic
models of SDSS~J1426+5752, how is it possible then that a white dwarf
with such a  high magnetic field, and perhaps no significant convection
zone, pulsates? 

If we take a look across the wide field of stellar oscillations, we
find that there is a perfect example of a class of objects where
pulsations and magnetism are found to coexist: the rapidly oscillating
Ap (roAp) stars \citep{kurtz2006}. These stars have a strong (by main
sequence standards) magnetic field (1000-10,000 G), no convection, and
pulsation modes are excited through a kappa-type mechanism. It could thus
be that SDSS~J1426+5752 is a white dwarf analog of the roAp stars. 
Amusingly, it is also not impossible that it might have pulsated as a
roAp itself in a distant past! The full nonadiabatic calculations 
presented in \citet{dufour08} rely on equilibrium models that all have
convection zones, and are thus likely inappropriate for the case of
SDSS~J1426+5752. One interesting avenue we intend to explore is to
construct models in which we would prevent artificially convection in
order to mimic the effects of the magnetic field. Would artificially
radiative models pulsate? We do not know yet, but it would not be
surprising that they could via the usual kappa-mechanism since there
would still be, convection or not, a huge opacity peak in the envelope
of such models. While the magnetic field is probably sufficiently strong
to inhibit convective motions, it is not strong enough to stop the
pulsations themselves, very obviously because we detect oscillation
modes. The effect of the strong field on the pulsations is probably
indirect in that it changes the conditions for driving, but much more
work is required before we understand exactly how this occurs. The
presence of the field may also force the pulsations to align themselves
on the magnetic axis as in roAp stars. The body of knowledge gathered so
far on these stars should be extremely useful as a guide for future
investigations of the pulsation properties of SDSS~J1426+5752. 

\section{CONCLUSION}\label{conclusion}

We presented evidence that SDSS~J1426+5752 is the first pulsating white
dwarf with the clear presence of a strong (i.e., strong enough for Zeeman
splitting to be observed) magnetic field ($\approx$ 1.2 MG). Such a
strong magnetic field probably inhibits the convective motions in this
object, but it is unclear yet how the pulsations are affected. We
proposed that this strange object could be a white dwarf analog of the
rapidly oscillating Ap stars and that a usual kappa-mechanism, even
in the absence of convection, might still explain the pulsational
instabilities. The confirmed presence of helium in the
envelope/atmosphere of SDSS~J1426+5752 is likely to play a key role in
this process. Future work testing this hypothesis is underway.

\acknowledgements

The authors would like to thank Constance Rockosi for her assistance
in obtaining the Keck Observations and Michael Bolte for facilitating
these observations. P.D. acknowledges the financial support of
NSERC. This work was also supported by the NSF through grant AST
03-07321 and AST-0602288. Support from the Reardon Foundation is also
gratefully acknowledged. G.F. acknowledges the contribution of the
NSERC and the Canada Research Chair Program.

\end{document}